\documentclass[aps,prl,twocolumn,showpacs,amsmath,amssymb]{revtex4}
\usepackage{epsfig}
\usepackage{bm}

\newcommand{\be}{\begin{equation}} \newcommand{\ee}{\end{equation}}

\newcommand{\ba}{\begin{eqnarray}}
\newcommand{\ea}{\end{eqnarray}}

\newcommand{\bq}{\begin{equation}}
\newcommand{\eq}{\end{equation}}
\newcommand{\bqa}{\begin{eqnarray}}
\newcommand{\eqa}{\end{eqnarray}}
\newcommand{\ben}{\begin{enumerate}}
\newcommand{\een}{\end{enumerate}}
\newcommand{\bc}{\begin{center}}
\newcommand{\ec}{\end{center}}
\newcommand{\bqb}{\begin{eqnarray*}}
\newcommand{\eqb}{\end{eqnarray*}}

\begin{document}

\title{\vspace{1cm}
Testing soft electroweak SUSY breaking from 
neutralino, chargino, and charged Higgs boson pairs production
at linear colliders
\footnote{Partially supported by EU contract HPRN-CT-2000-00149}}

\author{M. Beccaria$^{a,b}$, H. Eberl$^c$, 
F.M. Renard$^d$ and C. Verzegnassi$^{e, f}$ 
}

\affiliation{
$^a$Dipartimento di Fisica, Universit\`a di
Lecce \\
Via Arnesano, 73100 Lecce, Italy.\\
$^b$INFN, Sezione di Lecce\\
$^c$Institut f\"ur Hochenergiephysik der \"Osterreichischen Akademie
der Wissenschaften,\\ A-1050 Vienna, Austria\\
$^d$ Physique
Math\'{e}matique et Th\'{e}orique, UMR 5825\\
Universit\'{e} Montpellier
II,  F-34095 Montpellier Cedex 5.\hspace{2.2cm}\\
$^e$
Dipartimento di Fisica Teorica, Universit\`a di Trieste, \\
Strada Costiera
 14, Miramare (Trieste) \\
$^f$ INFN, Sezione di Trieste\\
}

\begin{abstract}
We consider the production of neutralino, chargino,
and charged Higgs boson pairs in the MSSM framework at future
$e^+e^-$ colliders. We show that, for c.m. energies in the one TeV
range and in a ``moderately'' light SUSY scenario, a combined
analysis of the slopes of these production cross sections could
lead to a strong consistency test involving the 
soft supersymmetric breaking parameters $M_1$, $M_2$, the Higgsino
parameter $\mu$, and $\tan\beta$.
\end{abstract}


\maketitle

It is a widespread hope within the members of the high energy elementary
particle physics community that the upcoming experiments at 
Tevatron~\cite{Tevatron} and at LHC~\cite{LHC} 
will finally reveal the existence of supersymmetric particles, via direct
production of sparticle-antisparticle pairs. In such an exciting hypothesis, the
simplest available theoretical supersymmetric model, the MSSM, will acquire 
the same role that belonged to the SM after the W, Z discoveries, and a long and
crucial period of precision tests will start, aiming to confirm, or to disprove,
the main theoretical assumptions that were used in the practical construction 
of the model.

For the MSSM case,
this program requires the construction of a linear $e^+e^-$ collider~\cite{LC}
that explores the c.m. TeV energy region. This machine should be sufficiently accurate to 
provide the same kind of consistency tests of the model that were achieved in the
hundred GeV region at LEP for the Standard Model, from detailed analyses of several
independent one-loop virtual effects.

If such an ambitious project is to be carried on, a clean investigation strategy
will undoubtedly be welcome, given the fact that the involved theoretical model 
has a not simple structure in which several independent parameters are 
implied. As a consequence of this feature, it might be appreciable
to identify clean experimental measurements that, possibly, isolate in a 
relatively simple way a reduced subset of parameters and of theoretical 
assumptions behind them.

The aim of this short paper is to show that a dedicated combined
analysis of the process of production of neutralino, chargino and charged
Higgs boson pairs might lead to a relevant precision test of the theoretical
assumptions that were used to fix, not only the basic supersymmetric
sector (i.e. the chiral superpotential and the gauge multiplet), but also the 
gaugino component of the soft supersymmetry breaking of the model. This will be
done in the paper by assuming a scenario of ``moderately'' light SUSY particles
(i.e. the various masses are all below, roughly, 350 GeV) and a 
c.m. energy $\sqrt{s}$ in the one TeV region. 
Under these two assumptions, a simple asymptotic expansion of so called
``Sudakov type'' is effective~\cite{Melles}. But the approach would not
change for larger SUSY masses $< 700$ GeV since a related c.m. energy 
rescaling would still be possible, from our previous experience, up to 
$\sqrt{s}$ values of about 2 TeV, where the resummation to higher orders
is still not necessary~\cite{Melles}.

After this brief description of the strategy of our approach
we are now ready to illustrate the practical details of our work. With this
aim, we must recall, for the reader's convenience, that the relevant
details of an analogous procedure, limited to the combination of chargino and 
charged Higgs pairs production, have already appeared in a previous
paper\cite{CharginoPaper}. It was shown there that from the combined
analyses of the slopes of the total cross sections in a c.m. energy range of 
about 1~TeV, assuming a ``light'' SUSY scenario where all the relevant 
sparticle masses lie below $\sim$ 350-400 GeV , a strong constraint on $\tan\beta$
(mostly produced by the Higgs cross section) and a strip in the $(M_2, \mu)$
plane (only produced by the chargino cross sections) were derivable. Here
$M_2$ is the soft SUSY breaking wino mass, $\mu$ is the Higgsino mass parameter,
and $\tan\beta=v_2/v_1$ is the 
ratio of the two Higgs doublets vevs. The novel process that we  
consider in this paper is neutralino pair production, and we shall 
show that its addition to the other processes will lead to highly improved 
constraints on the previous parameters, with the additional presence of $M_1$, the 
soft SUSY breaking bino mass. With this aim, we now briefly list the 
relevant theoretical formulae that we need for the analysis.

At the Born level, the neutralino pair production process 
$e^+e^-\to\chi^0_i\chi^0_j$ is described by 
three s, t, u channel components of the scattering amplitude, using the
following notations (the indices $a$, $b$ label the electron and the $i$-th 
neutralino chiralities):
\be
A^{ab}_{ij}\equiv{e^2\over s}S^{ab}_{ij}+{e^2\over u}U^{ab}_{ij}
+{e^2\over t}T^{ab}_{ij}
\ee
with the tree level values for the s channel 
\bq
\label{BornAmplitudeStart}
S^{LL}_{ij}={2s^2_{\rm w}-1\over4s^2_{\rm w}c^2_{\rm w}}O_{ij}
~~~~~~~
S^{LR}_{ij}=-~{2s^2_{\rm w}-1\over4s^2_{\rm w}c^2_{\rm w}}O^*_{ij}
\eq
\bq
S^{RL}_{ij}={1\over2c^2_{\rm w}}O_{ij}
~~~~~~~
S^{RR}_{ij}=-{1\over2c^2_{\rm w}}O^*_{ij}
\eq
\bq
O_{ij} = Z^{N*}_{3i}Z^{N}_{3j}-Z^{N*}_{4i}Z^{N}_{4j}
\eq
\noindent
and for the t, u channels
\bq
U^{LL}_{ij}=-~{1\over4s^2_{\rm w}c^2_{\rm w}}(Z^{N*}_{1i}s_{\rm w}
+Z^{N*}_{2i}c_{\rm w})(Z^N_{1j}s_{\rm w}
+Z^N_{2j}c_{\rm w})
\eq
\bq
U^{RR}_{ij}=-~{1\over c^2_{\rm w}}Z^{N}_{1i}Z^{N*}_{1j}
\eq
\bq
T^{LR}_{ij}={1\over4s^2_{\rm w}c^2_{\rm w}}(Z^{N*}_{1j}s_{\rm w}
+Z^{N*}_{2j}c_{\rm w})(Z^N_{1i}s_{\rm w}
+Z^N_{2i}c_{\rm w})
\eq
\bq
\label{BornAmplitudeEnd}
T^{RL}_{ij}={1\over c^2_{\rm w}}Z^{N}_{1j}Z^{N*}_{1i}
\eq
The quantities $Z^N_{ij}$ are the elements of the 
4x4 mixing matrix, defined in a conventional way~\cite{Melles}.
For simplicity, we 
have neglected in the considered asymptotic
region the selectron mass in Eq.(1).

For the purposes of a precision test, it becomes mandatory to compute the 
following perturbative one-loop expansion of the scattering amplitude. This 
requires the  calculation of a large number of Feynman diagrams. The 
resulting expressions are valid for any value of the c.m. energy 
$\sqrt{s}=\sqrt{(p_{e^-}+p_{e^+})^2}$. The full one-loop
result has been calculated recently~\cite{Helmut}.
In this work we propose an alternative approach, valid for values
of $\sqrt{s}$ ``much'' larger than all the SUSY
sparticle masses involved in the process. In fact, our strategy should be now
made very clear. We assume a previous production of 
the charginos, of the charged Higgs bosons and of at
least two  neutralinos with mass $M_{\chi^0_1}, M_{\chi^0_2}$.
Calling $M_{\rm SUSY}$ the heaviest of 
the real and virtual SUSY particles that appear in the processes,
and assuming a ``reasonable'' limit $M_{\rm SUSY}\simeq 350-400$ GeV, we
shall choose the value $\sqrt{s} \simeq 1$ TeV value to proceed with 
our approach. This is due to the fact that, in a previous paper only 
concerned with the charged Higgs pair production~\cite{ChargedHiggs},
we proved that in such a configuration an asymptotic energy
logarithmic expansion of so called ``Sudakov type'' was reliable,
with the only addition of a constant term to the leading quadratic and 
next-to-leading linear logarithmic terms. This conclusion allows
to propose a determination of the SUSY parameters entering the
Sudakov logarithms, based on measurements of the slope of the total
cross section, in which the (complicated) constant terms cancel.

The approach that we shall follow in this paper, that was already used in the 
combined chargino-charged Higgs analysis~\cite{CharginoPaper}, will \underline{assume}
that a similar expansion is valid, with only logarithmic
and constant terms. To prove this statement would require a detailed
comparison of the complete existing calculation~\cite{Helmut}
with the assumed asymptotic expansion. This will be the goal
of a forthcoming rigorous analysis. For the moment,
we shall assume its validity as a working Ansatz, and we shall show the
main relevant consequences that it will be able to produce.

After this preliminary discussion, we are now ready to write
the relevant asymptotic expansion of the scattering amplitude at one
loop, in which we shall only retain the leading quadratic and next-to-leading
linear logarithmic terms. For this purpose, we shall use the 
following formal decomposition the separate 
$F=S$, $T$ or $U$ subamplitudes (the Born values are those listed in 
Eqs.~(\ref{BornAmplitudeStart}-\ref{BornAmplitudeEnd})): 
\ba
F^{ab}_{ij} &=& F^{ab,~Born}_{ij}
+F^{ab,~Born}_{ij} c^{in}_{a}
+F^{ab,fin}_{ij} + \nonumber\\
&+& F^{ab,~ang}_{ij}+F^{ab,~RG}_{ij}
\ea
In the asymptotic expansion, two different kinds of logarithmic terms
appear. The first ones are the standard linear ones of RG origin. They
are well known and can be derived in a straightforward way 
replacing in the Born quantities the various bare
couplings  with running ones. For sake of completeness, 
we write the related formulae, noticing that they are only requested for the 
s-channel (purely 
Higgsino) component, as:
\ba
\label{initial}
S^{LL,RG}_{ij}&=&-~{\alpha\over4\pi}[\log{s\over M^2}]O_{ij}
({\tilde \beta'_0\over c^4_{\rm w}}-{\tilde\beta_0\over s^4_{\rm w}})\\
S^{LR,RG}_{ij}&=&{\alpha\over4\pi}[\log{s\over M^2}]O^*_{ij}
({\tilde \beta'_0\over c^4_{\rm w}}-{\tilde\beta_0\over s^4_{\rm w}})\\
S^{RL,RG}_{ij}&=&-~{\alpha\over\pi}[\log{s\over M^2}]O_{ij}
({\tilde \beta'_0\over 2c^4_{\rm w}})\\
S^{RR,RG}_{ij}&=&~{\alpha\over\pi}[\log{s\over M^2}]O^*_{ij}
({\tilde \beta'_0\over 2c^4_{\rm w}})
\ea
where ${\tilde \beta}_0= -1/4$ and ${\tilde \beta}_0^\prime = -11/4$.
The second type of logarithms which arise asymptotically is that of the 
``genuine weak'' Sudakov terms. These are usually classified~\cite{SudakovClass}
as logarithms of gauge non universal, gauge universal and Yukawa origin.
The gauge non universal, scattering angle dependent ones, stem from box
diagrams and t,u  channel vertices and have the following 
expressions:
\ba
S^{LL,ang}_{ij}&=&-~{\alpha\over4\pi s^4_{\rm w}}
[\log{s\over M^2_{\rm w}}][\log\frac{-t}{s}
[2Z^{N*}_{2i}Z^{N}_{2j}+Z^{N*}_{4i}Z^{N}_{4j}]\nonumber\\
&& -\log\frac{-u}{s}
[2Z^{N}_{2j}Z^{N*}_{2i}+Z^{N}_{3j}Z^{N*}_{3i}]] 
\ea
\ba
S^{LR,ang}_{ij}&=&-~{\alpha\over4\pi s^4_{\rm w}}
[\log{s\over M^2_{\rm w}}][\log\frac{-t}{s}
[2Z^{N}_{2i}Z^{N*}_{2j}+Z^{N}_{3i}Z^{N*}_{3j}] \nonumber\\
&&-\log\frac{-u}{s}[2Z^{N*}_{2j}Z^{N}_{2i}+Z^{N*}_{4j}Z^{N}_{4i}]]
\ea
\bqa
U^{LL,ang}_{ij}&=&~{\alpha\over4\pi s^4_{\rm w}c_{\rm w}}
[\log{s\over M^2_{\rm w}}]
\{[Z^{N*}_{2i}(Z^{N}_{1j}s_W-Z^{N}_{2j}c_W) \nonumber \\
&& +Z^{N}_{2j}(Z^{N*}_{1i}s_W-Z^{N*}_{2i}c_W)]\log\frac{-t}{s}
\nonumber\\
&&
+[Z^{N*}_{2i}(Z^{N}_{1j}s_W+Z^{N}_{2j}c_W) \nonumber \\
&&
+Z^{N}_{2j}(Z^{N*}_{1i}s_W+Z^{N*}_{2i}c_W)]\log\frac{-u}{s}]\}
\eqa
\bqa
T^{LR,ang}_{ij}&=&-~{\alpha\over4\pi s^4_{\rm w}c_{\rm w}}
[\log{s\over M^2_{\rm w}}]
\{[Z^{N*}_{2j}(Z^{N}_{1i}s_W-Z^{N}_{2i}c_W) \nonumber\\
&&
+Z^{N}_{2i}(Z^{N*}_{1j}s_W-Z^{N*}_{2j}c_W)]\log\frac{-u}{s}
\nonumber\\
&&
+[Z^{N*}_{2j}(Z^{N}_{1i}s_W+Z^{N}_{2i}c_W)\nonumber \\
&&
+Z^{N}_{2i}(Z^{N*}_{1j}s_W+Z^{N*}_{2j}c_W)]\log\frac{-t}{s}]\}
\eqa
The gauge universal ones are due both to the initial
and to the final vertices. The initial contribution is 
fixed by the coefficients
\bq
c^{in}_L={\alpha(1+2c^2_W)\over16\pi s^2_W c^2_W}
[~2\log{s\over M^2_{\rm w}}-\log^2{s\over M^2_{\rm w}}~]
\eq
\bq
c^{in}_R={\alpha\over4\pi c^2_W}
[~2\log{s\over M^2_{\rm w}}-\log^2{s\over M^2_{\rm w}}~]
\eq
The contribution from final neutralino legs is different for the 
s or u, t channels. For the s-channel we have higgsino components
\bqa
S^{ab,\rm fin}_{ij}&=& \sum_k \{
S^{ab,\rm Born}_{ik}. [c^{\rm fin~gauge}_{kj}
+c^{\rm fin~Yukawa}_{kj}]+\nonumber\\
&+&
S^{ab,\rm Born}_{kj}. [c^{\rm fin~gauge}_{ki}
+c^{\rm fin~Yukawa}_{ki}]^*\}
\eqa
where
\bqa
\label{fingauge}
\lefteqn{c^{\rm fin~gauge}_{kj} = ({\alpha\over\pi})
{(1+2c^2_{\rm w})\over 32s^2_{\rm w}c^2_{\rm w}}
~[~2\log{s\over M^2_{\rm w}}-\log^2{s\over M^2_{\rm w}}~]} &&
\\
&& [(Z^{N*}_{4k}Z^{N}_{4j}+Z^{N*}_{3k}Z^{N}_{3j})P_L
+(Z^{N}_{4k}Z^{N*}_{4j}+Z^{N}_{3k}Z^{N*}_{3j})P_R]\nonumber
\eqa
The logarithms of Yukawa origin are only due to 
final s-channel vertices with virtual heavy (b,t) quark-squark pairs.
These are the only logarithmic terms that contain (also) the vevs ratio $\tan\beta$
and their expression is 
\bqa
c^{\rm fin~Yukawa}_{kj}&=&({\alpha\over\pi})
[-\log{s\over M^2}]~({3\over 16s^2_{\rm w} M^2_{\rm w}})~[
\\
&&{m^2_t\over \sin^2\beta}
~(Z^{N*}_{4k}Z^{N}_{4j}P_L+Z^{N}_{4k}Z^{N*}_{4j}P_R)\nonumber\\
&&
+{m^2_b\over \cos^2\beta}
~(Z^{N*}_{3k}Z^{N}_{3j}P_L+Z^{N}_{3k}Z^{N*}_{3j}P_R)] \nonumber
\eqa
To conclude, the final contribution to the u and t channel 
amplitudes is due to gaugino components for which there is no 
Yukawa contribution:
\bqa
U^{ab,\rm fin}_{ij}=\sum_k \{U^{ab,\rm Born}_{ik}.[c^{\rm fin~gauge}_{kj}]
+U^{ab,\rm Born}_{kj}.[c^{\rm fin~gauge}_{ki}]^*\}\eqa
\bqa
T^{ab,\rm fin}_{ij}=\sum_k \{T^{ab,\rm Born}_{ik}.[c^{\rm fin~gauge}_{kj}]
+T^{ab,\rm Born}_{kj}.[c^{\rm fin~gauge}_{ki}]^*\}
\eqa
with 
\ba
c^{\rm fin~gauge}_{kj}&=& -\frac{\alpha}{4\pi\ s_{\rm w}^2}
[Z^{N*}_{2k}Z^N_{2j}P_L+Z^{N}_{2k}Z^{N*}_{2j}P_R] \nonumber\\
\label{final}
&& \times \log^2\frac{s}{M_V^2}.
\ea

Eqs.(\ref{initial}-\ref{final}) 
represent the complete logarithmic contributions at one loop
to the considered process, and are the main original result of this paper.
We expect from our previous discussion their reliability for what concerns
the calculation of the slope of the total cross section in the 1 TeV range. To
perform the latter, one must first compute the expression of the differential
cross section that is readily obtained from the above amplitude.
After angular integration, one obtains the \underline{approximate}
(to next-to-leading logarithmic order) asymptotic expression of the 
total cross section. This will be sufficient to determine the
\underline{effective} asymptotic expression of the slope of the 
cross section, where by definition possible extra ``next-to-next-to-leading''
(i.e. constant) terms disappear. This expression will only depend on $\tan\beta$
and on the mixing parameters $Z_{ij}^N$. The latter ones, in turn, can be expressed 
as functions of
the supersymmetric parameters $M_1$, $M_2$, $\mu$ and $\tan\beta$. Thus, the final form of the 
slope will depend on these four parameters.

In order to achieve the maximal theoretical information using our approach, we have combined the 
theoretical expressions of the slope of the neutralino pair total cross section, given
in this paper, with those of the chargino and charged Higgs pairs, given 
in~\cite{CharginoPaper}. In practice, we have limited our analysis to the 
production of the two light observable final states ($\chi_1^0\chi_2^0$, 
$\chi_2^0\chi_2^0$) for neutralinos, combined with the production of all
the three chargino pairs and of the charged Higgs pair. 
To work in a consistent way, we have chosen as relevant examples for the masses of the produced 
pairs three sets of values denoted by $S_1$, $S_2$, and $S_3$. 
The first, $S_1$, is the Tesla benchmark point  
RR2\cite{RR2} with the two lightest neutralinos being respectively 95 \% bino 
and 82 \% wino. The set $S_2$ is a mixed scenario with neutralinos having non negligible
gaugino and Higgsino components; $\chi^0_1$ is 86 \% bino and 13 \% higgsino, 
$\chi^0_2$ is 11 \% bino, 48 \% wino and 41 \% higgsino. 
Finally, $S_3$ is a purely Higgsino one  with the 
two lightest neutralinos being 92 \% and 98 \% higgsino like.
The values of the input parameters as well as the masses of the two charginos and of 
the two lightest neutralinos are summarized in Tab.~(1).
We have performed a standard $\chi^2$ analysis assuming in the various 
scenarios 10-12 experimental points at energies ranging from 700-850 GeV 
(depending on the scenario) up to 1200 GeV and assuming an experimental accuracy of 
1\% for the cross sections of gauginos and 2\% for that of Higgs bosons.
For each scenario we have checked the condition $\sqrt{s}\gg M_{\rm SUSY}$. 
The results of our analysis are shown in Figs.(1-3).
From inspection of the figures one can draw
the following conclusions. 
\begin{table}
\caption{\label{tab:table1}Input parameters and masses of charginos, and lightest neutralinos 
for the three input sets $S_1$, $S_2$, and $S_3$. The masses are expressed in GeV.}
\begin{ruledtabular}
\begin{tabular}{lccc}
& $S_1$ & $S_2$ & $S_3$ \\
\hline
$M_1$        &  78 & 100 & 200\\
$M_2$        & 150 & 200 & 400\\
$\mu$        & 263 & 200 & 100\\
$\tan\beta$  &  30 &  30 &  30\\
\hline
$\chi^\pm_1$ & 132 & 149 &  95\\
$\chi^\pm_2$ & 295 & 266 & 417\\
$\chi^0_1$   &  75 &  92 &  82\\
$\chi^0_2$   & 133 & 153 & 109
\end{tabular}
\end{ruledtabular}
\end{table}
\begin{figure}[htb]
\vskip 0.4cm
\begin{center}
\leavevmode
\psfig{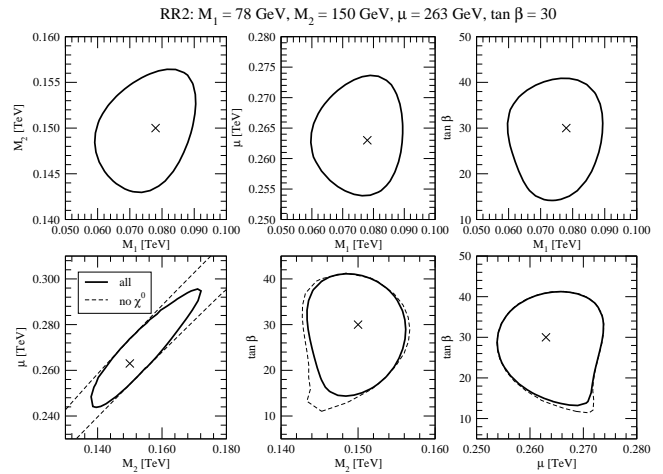}
\caption{$S_1$ benchmark point. $1\sigma$ error bounds on the MSSM parameters $M_2$, $\mu$, $\tan\beta$.
In this and in the following figures the crosses denote the values of the parameters in the 
specific benchmark.}
\label{fig:RR2}
\end{center}
%\vspace{-5mm}
\end{figure}
\noindent
In Fig.(1) we consider the $S_1$ benchmark point. One notices that a 
combined analysis is able to
generate closed contours in the planes of all possible six couples of
parameters. This possibility is obviously existing for the first 
3 cases i.e. $(M_1, M_2)$, $(M_1, \mu)$, $(M_1, \tan\beta)$ under the 
condition that the neutralino information is added to the remaining ones, that do not
depend on $M_1$. Less obvious and more illustrative is the closure of the 
undetermined strip in the $(M_2, \mu)$ plane (drawn in the absence of 
neutralino data) when neutralino data are added to the remaining 
ones. This addition does not practically improve, on the contrary, the bounds on 
$(M_2, \tan\beta)$ and $(\mu, \tan\beta)$ obtained from chargino-Higgs data.
One sees typical errors, under the assumed experimental conditions, of about
10 GeV for $M_1$, $M_2$, $\mu$ and of a relative $\sim 30-40$ \%
for $\tan\beta$.
\begin{figure}[htb]
\vskip 0.4cm
\begin{center}
\leavevmode
\psfig{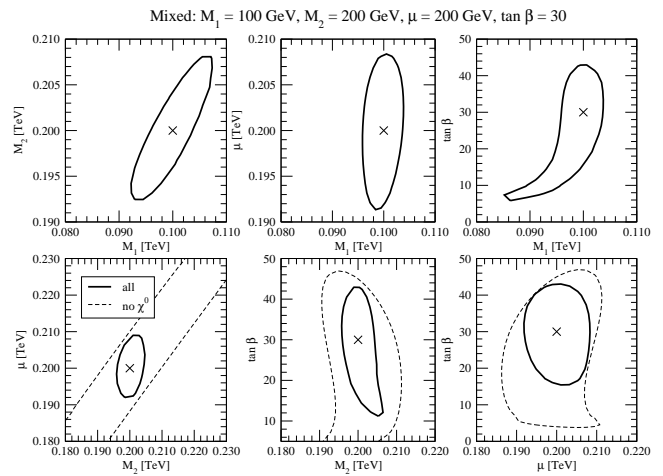}
\caption{$S_2$ benchmark point. $1\sigma$ error bounds on the MSSM parameters $M_2$, $\mu$, $\tan\beta$.}
\label{fig:SPS4}
\end{center}
\vspace{-5mm}
\end{figure}
\noindent
In Fig.(2) we consider the $S_2$ scenario. 
One finds, approximately, the same features for the set of 
projections. The size of errors is now 
of approximately $\sim 10$ GeV for $M_1$, $M_2$, and $\mu$, a relative 40-50 \%
for $\tan\beta$.
\begin{figure}[htb]
\begin{center}
\leavevmode
\psfig{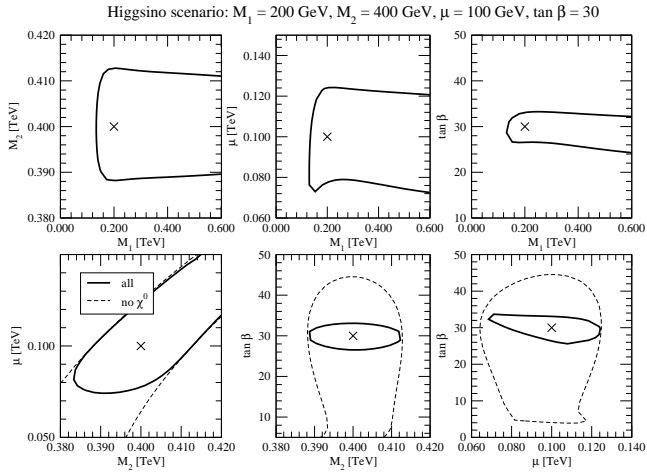}
\caption{$S_3$ scenario. 
$1\sigma$ error bounds on the MSSM parameters $M_2$, $\mu$, $\tan\beta$.}
\label{fig:Higgsino}
\end{center}
\vspace{-5mm}
\end{figure}
\noindent
Fig.(3) describes $S_3$, a typical Higgsino scenario.
Here the new (negative) feature concerns the lack of 
boundary for $M_1$, as seen from the 
first three plots. This can be understood since, at the considered 
$M_1$, $M_2$ and $\mu$ values, the mixing parameters $Z_{ij}^N$ do not vary appreciably
when $M_1$ increases. On the contrary, the three plots in the bottom line
retain the typical features of the previous cases, with particular
relevance of the neutralino role for the determination of lower limits for $\mu$ and $M_2$. The 
errors are now , typically, of about 20 GeV for $M_2$, $\mu$ and of a relative $\sim 10\%$
for $\tan\beta$.

The previous results should illustrate the aimed possible outcomes
of the testing strategy that we are proposing in this preliminary qualitative
paper. Our assumed physical inputs will necessarily be the masses of the 
(supposedly produced) (light) neutralinos, charginos and charged Higgs. Their values will
depend on the four parameters $M_1$, $M_2$, $\mu$, $\tan\beta$ but, in general
there will not be a 1-1 correspondence, and different sets of parameters might
reproduce ``essentially'' the same masses~\cite{MassesVsParameters}.
Within the errors that we have illustrated in our examples, these
different sets can be discriminated via the $\chi^2$ analysis that we have proposed,
and the ``correct'' set is selected by the true experimental data, 
if the latter are actually described by the model. If none of the candidate 
sets were compatible with the analysis, an indication would arise that some of the 
details of the model might need a suitable modification.

\end{document}